\documentclass [preprint]{revtex4}
\usepackage{epsfig}
\usepackage{graphicx}
\usepackage{amsmath,amscd,amssymb,graphicx,verbatim}
\newtheorem{theorem}{Theorem}
\newcommand{\fl}{}
\begin{document}

\title
{A  nontrivial bosonic representation of large spin systems  at high temperatures}
\author{Yamen  Hamdouni}\thanks{ Physics departement, Mentouri University-Constantine 1}
\email{hamdouniyamen@gmail.com}

\begin{abstract}
We report on a nontrivial bosonization scheme for spin operators. It is shown that in the large $N$ limit, at infinite temperature, the operators $\sum_{k=1}^N \hat s_{k\pm}/\sqrt{N}$ behave like the creation and annihilation operators, $a^\dag$ and $a$, corresponding to a harmonic oscillator in thermal equilibrium, whose temperature and frequency are related by $\hbar\omega/k_B T=\ln 3$. The $z$ component is found to be equivalent to the position variable of another harmonic oscillator occupying its ground Gaussian state at zero temperature. The obtained results are applied to the Heisenberg $XY$ Hamiltonian at finite temperature.    
\end{abstract}

\pacs{03.65.-w, 05.30.-d} 
\maketitle
\section{Introduction}
The interest in bosonic representations of spin operators arises from the necessity of finding alternative practical tools to the diagrammatic technique for spins which turns out to be quite complicated~\cite{1}. By using bosonic operators, one can, for instance, evaluate the functional integrals and afterwards calculate the different contributions diagrammatically~\cite{2,3,4}.
 Moreover, spin-wave theory is based on the possibility of transforming spin operators into new bosonic operators, which are much more easier to deal with~\cite{5}. This theory has known great success in the study of magnetic properties of many materials at low temperatures; it led, in particular, to the notion of magnon, the elementary collective excitation of spins in the crystal lattice. Many important spin Hamiltonians, such as the Heisenberg model Hamiltonian, can be exactly diagonalized at low excitations with respect to the bosonic degrees of freedom  using certain transformations.

 Among the most used transformations  in spin-wave theory we cite the Dyson~\cite{6} and the Holstein-Primakoff transformations. The latter maps spin-$\frac{1}{2}$ operators as follows~\cite{7}:
\begin{eqnarray}
\hat s_+&=&a^\dag \sqrt{1-a^\dag a} , \label{hol1}\\ 
\hat s_-&=&\sqrt{1-a^\dag a} a,\label{hol2} \\
\hat s_z&=&-\frac{1}{2}+a^\dag a. \label{hol3}
\end{eqnarray}
Here and throughout the text, unless otherwise stated, we  take $\hbar$=1. For a set of $N$ spins, the  scaled total raising operator is thus given by
\begin{equation}
\frac{\hat S_+}{\sqrt{N}}=\frac{1}{\sqrt{N}}\sum_{k=1}^{N} a_k^\dag\sqrt{1-a_k^\dag a_k} .
\end{equation} 
At low temperatures and low excitations, i.e. when $\langle a_k^\dag a_k\rangle<<1$, we may, as a first approximation, write
\begin{equation}
\frac{\hat S_+}{\sqrt{N}}=\frac{1}{\sqrt{N}}\sum_{k=1}^{N} a_k^\dag.
\end{equation} 
For large $N$, the Fourier transforms of the bosonic operators $a_k$ and $a^\dag_k$ are given by
\begin{equation}
\alpha_\ell=\frac{1}{\sqrt{N}}\sum_k^N e^{-i k\ell} a_k, \qquad  \alpha_\ell^\dag=\frac{1}{\sqrt{N}}\sum_k^N e^{i k\ell} a_k^\dag.
\end{equation}
These are also bosonic operators, since they verify the commutation relation $[\alpha_\ell,\alpha_{\ell'}^\dag]=\delta_{\ell\ell'}$. We can thus conclude that at low excitations (temperatures), and large $N$, 
\begin{equation}
 \frac{\hat S_+}{\sqrt{N}}=\alpha_0^\dag,\quad \frac{\hat S_-}{\sqrt{N}}=\alpha_0\label{nova}
\end{equation}
which is indeed a beautiful result. However, at higher excitations, the square root in equations~(\ref{hol1})-(\ref{hol3}) should be expanded in powers of $a^\dag a$, depending on the order of the adopted approximation. This gives, for instance,
\begin{equation}
 \frac{\hat S_+}{\sqrt{N}}=\frac{1}{\sqrt{N}}\sum_{k=1}^N a_k^\dag(1-\frac{a_k^\dag a_k}{2}-\frac{(a_k^\dag a_k)^2}{8}-\frac{(a_k^\dag a_k)^3}{16}-\cdots) .
\end{equation}

In this paper we  address the problem of finding a suitable bosonic representation similar to~(\ref{nova}) that is valid at high temperatures. As we shall see, the cost we have to pay for preserving the above form of the transformation resides in the introduction of effective temperatures that are generally different from the actual ones, together with a deformation of the functions of interest.

 The paper is organized as follows. In Section~\ref{sec2}
we prove some fundamental results in connection with the addition of spin operators. The focus is on the trace properties of the  suitably scaled operators to ensure good statistical behaviour in the large $N$ limit. Our main result is presented in Section~\ref{sec3} where we explicitly establish the exact form of the mapping between the spin and bosonic operators. The paper is ended with a brief discussion.

\section{Preliminary results\label{sec2}} 

Let us begin by proving some results about the behaviour of the spin operators $\hat S_\alpha/\sqrt{N}=\sum_{i=1}^N\hat s_{i\alpha}/\sqrt{N}$ ($\alpha\equiv x, y,z$) as $N\to\infty$~\cite{8,9,10,11,12}. By the multinomial theorem we have that
\begin{equation}
{\rm tr}\bigl( \hat S_\alpha\bigr)^\ell=\sum_{r_1 r_2\cdots r_N}\frac{\ell!}{r_1! r_2!\cdots r_N!}{\rm tr} \hat s_{1\alpha }^{r_1}\otimes \hat s_{2\alpha}^{\rm r_2}\otimes\cdots\otimes \hat s_{N\alpha}^{r_N}\delta\biggl(\ell-\sum_k r_k\biggr).
\end{equation}
Note that we are using the Kronecker delta symbol. We have to distinguish between two possible cases, depending on whether $\ell$ is odd or even. In the former case, we see that there should always be an operator, in each term of the right-hand side of the above equation, appearing with an odd power, say, $2n+1$ where $n\in \mathbb Z_+$. But since ${\rm tr} \hat s^{2n+1}_{  \alpha}=0$ for all $n$,  we conclude that
\begin{equation}
{\rm tr} \Biggl(\frac{\hat S_\alpha}{\sqrt{N}}\Biggr)^{2\ell+1}=0, \qquad \forall \ell\in \mathbb Z_+.
\end{equation}
When the power of the operator is even, 
\begin{equation}
{\rm tr}\bigl(\hat S_\alpha\bigr)^{2\ell}=\sum_{r_1 r_2\cdots r_N}\frac{2\ell!}{r_1! r_2!\cdots r_N!}\prod_{k=1}^N{\rm tr} \hat s_{k\alpha }^{r_k}\delta\biggl(2\ell-\sum_k r_k\biggr).\label{prod2}
\end{equation}  
One can easily see that the sum in the right-hand side of equation~(\ref{prod2}) yields a polynomial in $N$ of degree $\ell$, the main contribution of which comes from the term in which all the operators $\hat s_{k\alpha }$ contribute equally to the product. This happens when all the operators appear with the same power; all the other possibilities yield polynomials of degree less than $\ell$. But due to the restriction imposed on  the sum of the powers of the operators, we conclude that:
\begin{equation}
{\rm tr}\bigl(\hat S_\alpha\bigr)^{2\ell}=\frac{2\ell!}{\underbrace{2! \times 2!\cdots 2!}_{\ell \ \rm terms }}\sum_{\Pi_\ell} \Biggl(\prod_{k=1}^\ell{\rm tr}s^2_{\alpha k}\prod_{k=\ell+1}^N{\rm tr}{\mathbb I}_2\Biggr)_{\Pi_\ell[1,2,\cdots N]}+Q_{\ell-1}(N),
\end{equation}
 where the products are evaluated for all the possible partitions $\Pi_\ell[1,2,\cdots, N]$ of  $N$ elements into subsets  of $\ell$ elements, and $Q_{\ell-1}(N)$ is a polynomial in $N$ of degree at most equal to $\ell-1$.
Next we remark  that for spin-1/2 operators, $\hat s_\alpha^2=\frac{ \mathbb I_2}{4}$, meaning that ${\rm tr} \hat s_\alpha^2=\frac{1}{2}$, independent of the partition $\Pi_\ell$. Hence we can write
\begin{eqnarray}
\fl {\rm tr}\bigl(\hat S_\alpha\bigr)^{2\ell}&=&\frac{2\ell!}{2^{\ell}} \frac{N(N-1)(N-2)\cdots( N-\ell+1)}{\ell !} ({\rm tr\ s^2_\alpha})^\ell ({\rm tr}\ \mathbb I_2)^{N-\ell}+Q_{\ell-1}(N)\nonumber\\
\fl &=&2^N N^\ell \Biggl[\frac{(2\ell)!}{2^{3\ell}\ell!}\prod_{k=1}^{\ell-1}\Bigl(1-\frac{k}{N}\Bigr)+O\Biggr(\frac{1}{N}\Biggr)\Biggr].
\end{eqnarray}
It follows that
\begin{equation}
\lim_{N\to\infty} 2^{-N}{\rm tr} (\hat S_\alpha/\sqrt{N})^{2\ell}=\frac{(2\ell)!}{2^{3\ell}\ell!}, \qquad \alpha\equiv x, y, z.\label{moment}
\end{equation}
Now we shall prove that as $N\to\infty$ the operators $\hat S_\alpha/\sqrt{N}$ and $\hat S_\beta/\sqrt{N}$ become uncorrelated (i.e. independent) with respect to the tracial state when $\alpha\neq \beta$. We have
\begin{eqnarray}
\fl {\rm tr} \Big\{(\hat S_\alpha)^m (\hat S_\beta)^\ell\Bigr\}&=& \sum_{k_1,k_2,\cdots k_N, r_1, r_2\cdots r_N}\frac{m!\ell!}{r_1! r_2!\cdots r_N! k_1! k_2!\cdots k_N!}\nonumber\\ \fl
&\times& {\rm tr} \ \hat s_{1\alpha }^{r_1}\otimes \hat s_{2\alpha}^{\rm r_2}\otimes\cdots\otimes \hat s_{N\alpha}^{r_N} .  \ \hat s_{1\beta }^{k_1}\otimes \hat s_{2\beta}^{\rm k_2}\otimes\cdots\otimes \hat s_{N\beta}^{k_N} \nonumber \\
&\times& \delta\biggl(m-\sum_i r_i\biggr) \delta\biggl(\ell-\sum_i k_i\biggr)\nonumber\\
\fl &=& \sum_{k_1,k_2,\cdots k_N, r_1, r_2\cdots r_N}\frac{m!\ell!}{r_1! r_2!\cdots r_N! k_1! k_2!\cdots k_N!}\nonumber\\ \fl
&\times& {\rm tr} (\ \hat s_{1\alpha }^{r_1}. \ \hat s_{1\beta }^{k_1})\otimes (\hat s_{2\alpha}^{\rm r_2}.  \hat s_{2\beta}^{\rm k_2})\otimes\cdots\otimes (\hat s_{N\alpha}^{r_N}.\hat s_{N\beta}^{k_N}) \nonumber \\
&\times& \delta\biggl(m-\sum_i r_i\biggr) \delta\biggl(\ell-\sum_i k_i\biggr).\label{two}
\end{eqnarray}
Once again, if one of the powers is odd, then in each  term in the right-hand side of the latter equation there should be an operator $\hat s_j$ appearing with an odd power; this means that its trace is zero, and hence the overall trace vanishes as well. This gives 
\begin{equation}
{\rm tr}(\hat S_\alpha^{2\ell+1} \hat S_\beta^m)={\rm tr} (\hat S_\alpha)^{2\ell+1} {\rm tr} (\hat S_\beta)^m=0, \qquad \forall m\in \mathbb Z_+.
\end{equation} 
Consider now the case in which the powers of the total spin operators are  even and make  the substitution $m\rightarrow 2m$, $\ell\rightarrow 2\ell$ in equation~(\ref{two}). It follows that the right-hand side of the latter equation yields a polynomial in $N$ of degree $\ell+m$; we shall assume  that $\ell>m$. In this instance,  the leading contribution to the sum comes from the terms in which all the spins equally contribute to the trace, which occurs only if all their powers are equal to 2, taking, obviously, into account all the possible partitions of $N$ elements into disjoint subsets of $\ell$ and $m$ elements. Since we have assumed that $\ell>m$, we should have  a contribution to the trace of $m$ operators in the form $\hat s_{j\alpha}^2\hat s_{j\beta}^2$, and $\ell-m$ in the form $\hat s_{j\beta}^2$;  the  remaining $N-\ell$ traces  come from the identity matrix. More precisely: 
\begin{eqnarray}
\fl {\rm tr} \Big\{(\hat S_\alpha)^{2m} (\hat S_\beta)^{2\ell}\Bigr\}&=& \frac{(2m)!(2\ell)!}{\underbrace{2! 2!\cdots 2!}_{\ell \ {\rm terms}}\underbrace{ 2! 2!\cdots 2!}_{m \ {\rm terms}}}\sum_{\Pi_\ell\Pi_m}
 \Biggl(\prod_{j=1}^{m} {\rm tr}\ (\hat s_{ j\alpha}^2 \hat s_{ j \beta }^2) \prod_{j=1}^{\ell-m}  {\rm tr} \hat s_{j\beta}^2 \prod_{j=1}^{N-\ell}  {\rm tr} \ {\mathbb  I_2}\Biggr)_{\Pi_\ell, \Pi_m}\nonumber\\ \fl&+&P_{\ell+m-1}(N),
\end{eqnarray}
where $P_{\ell+m-1}(N)$ is a polynomial in $N$ of degree at most equal to $\ell+m-1$. Taking into account the fundamental properties of spin-$\frac{1}{2}$ operators, we find that 
\begin{eqnarray}
\fl {\rm tr} \Big\{(\hat S_\alpha)^{2m} (\hat S_\beta)^{2\ell}\Bigr\}&=&\frac{(2m)!(2\ell)!}{2^{\ell+m}}\times\frac{N(N-1)(N-2)\cdots(N-m+1)}{m!}\nonumber\\ \fl
&\times&\frac{N(N-1)(N-2)\cdots N(N-\ell+1)}{\ell!}\times\frac{2^{N-\ell}}{8^m 2^{\ell-m}}+P_{\ell+m-1}(N)\nonumber \\
&=& N^{\ell+m}\Biggl[\frac{2^N (2m)!(2\ell)!}{2^{3\ell}2^{3m} \ell! m!}\prod_{k=1}^{\ell-1}\Bigl(1-\frac{k}{N}\Bigr)\prod_{k'=1}^{m-1}\Bigl(1-\frac{k'}{N}\Bigr)+O\Bigl(\frac{1}{N}\Bigr)\Biggr].\label{down}
\end{eqnarray}
As a direct result of equation~(\ref{down}) we deduce that
\begin{equation}
\lim_{N\to\infty} 2^{-N}{\rm tr}\ \Big\{(\hat S_\alpha/\sqrt{N})^{2m} (\hat S_\beta/\sqrt{N})^{2\ell}\Bigr\}=\frac{ (2m)!(2\ell)!}{2^{3\ell}2^{3m} \ell! m!},
\end{equation}
which means that
\begin{equation}
\lim_{N\to\infty} 2^{-N}{\rm tr}\ \Big\{(\hat S_\alpha/\sqrt{N})^{2m} (\hat S_\beta/\sqrt{N})^{2\ell}\Bigr\}=\lim_{N\to\infty} 2^{-N}{\rm tr}\ (\hat S_\alpha/\sqrt{N})^{2m} \lim_{N\to\infty} 2^{-N}{\rm tr}\ (\hat S_\beta/\sqrt{N})^{2\ell}.
\end{equation}

Equation~(\ref{moment}) determines the explicit form of the moments of the random variable, $\eta$, associated with the spectral decomposition of the operators $\hat S_\alpha/\sqrt{N}$ as $N$ tends to infinity. To find the expression of the corresponding probability density function $p(\eta)$, we first note that the characteristic function is given in terms of the moments $\langle\eta^{n}\rangle$ by
\begin{equation}
\Phi(t)=\sum_{n=0}^\infty (i t)^n\frac{  \langle\eta^{n}\rangle}{n!},
\end{equation}
so that
\begin{equation}
\Phi(t)=\sum_{n=0}^\infty (it)^{2n} \frac{(2n)!}{2^{3n} n! (2n)!}=\sum_{n=0}^\infty \frac{ (it)^{2n}}{2^{3n} n! }.
\end{equation} 
 This yields
\begin{equation}
\Phi(t)=e^{-t^2/8}.
\end{equation}
The probability density function is nothing but the Fourier transform of the characteristic function:
\begin{equation}
p(\eta)=\frac{1}{2\pi}\int_{-\infty}^{\infty}\Phi(t) e^{-it \eta}dt.
\end{equation} 
By direct substitution of the expression of $\Phi(t)$ into the latter equation, we find that
\begin{equation}
p(\eta)=\sqrt{\frac{2}{\pi}}e^{-2\eta^2},
\end{equation}
that is, a Gaussian probability distribution with zero mean and variance $\sigma=\frac{1}{2}$.

A corollary of the above result is that for any well-behaved functions $f,g$:
\begin{equation}
\lim_{N\to\infty}2^{-N}{\rm tr} f\Bigl(\frac{\hat S_\alpha}{\sqrt{N}}\Bigr)=\sqrt{\frac{2}{\pi}}\int_{-\infty}^{\infty}f(\eta) e^{-2\eta^2} d\eta, \label{imad}
\end{equation}
\begin{equation}
\lim_{N\to\infty}2^{-N}{\rm tr} g\Bigl(\frac{\hat S_\alpha}{\sqrt{N}},\frac{\hat S_\beta}{\sqrt{N}}\Bigr)=\frac{2}{\pi}\iint\limits_{-\infty}^{+\infty}g(\eta,\mu)e^{-2(\eta^2+\mu^2)} d\eta d\mu,  \qquad \alpha, \beta\equiv x, y, z.\label{mass}
\end{equation}
If, instead of the operators $\hat S_x$ and $\hat S_y$, we use the raising  and lowering operators $\hat S_+$ and $\hat S_-$, which are related to each other by
\begin{equation}
\hat S_x=\frac{1}{2}(\hat S_++\hat S_-), \quad \hat S_y=\frac{1}{2i}(\hat S_+-\hat S_-),
\end{equation}
the integral in equation~(\ref{mass}) should be expressed in terms of the random variables $z^*$ and $z$ corresponding to $\hat S_+$ and $\hat S_-$, respectively. One can easily verify that
\begin{equation}
 \lim_{N\to\infty}2^{-N}{\rm tr} g\Bigl(\frac{\hat S_+}{\sqrt{N}},\frac{\hat S_-}{\sqrt{N}}\Bigr)=\frac{2}{\pi}\int_{z^*}\int_{z}g(z^*,z)e^{-2|z|^2} dz^* dz.\label{main1}
\end{equation}
This being explicitly proved, we are ready now to formulate our main result in the next section.
\section{Bosonization\label{sec3}}
The crucial observation is that equation~(\ref{main1}) is nothing but the coherent state representation of the mean value of the function operator $g$ written in normal order (which we denote by $\mathcal N g$): 
\begin{equation}
\frac{2}{\pi}\int_{z^*}\int_{z}g(z^*,z)e^{-2|z|^2} dz^* dz=\int_{z^*}\int_{z}P(z^*,z) {\mathcal N}g(z^*,z) dz^* dz.
\end{equation}
Here $P(z^*,z)$ is the P-representation~\cite{13} of the density matrix $\hat \rho$ of an equivalent bosonic system (harmonic oscillator, as we shall see bellow) whose creation and annihilation operators are denoted by $a^\dag$ and $a$, respectively. Explicitly, we have
\begin{equation}
P(z^*,z)=\sum_n\langle n|\hat\rho\delta(z^*-a^\dag)\delta(z-a)|n\rangle
\end{equation}
 which, obviously, is normalized to unity: 
\begin{equation}
\int_{z^*}\int_{z}P(z^*,z) dz^* dz=1.
\end{equation}
It follows that the density matrix $\hat\rho$ is given by
\begin{equation}
\hat\rho=\int_{z^*}\int_{z}P(z^*,z)|z\rangle \langle z|dz^* dz,
\end{equation}
with the coherent states
\begin{equation}
|z\rangle=e^{-|z|^2/2}\sum_{m=0}^{\infty}\frac{z^m}{m!}|m\rangle.
\end{equation}
Whence
\begin{eqnarray}
\langle n|\hat\rho|m\rangle&=&\frac{2}{\pi}\int_{z^*} dz^*\int_{z} dz \  e^{-3|z^2|}\frac{z^n{z^{*}}^m}{\sqrt{n!m!}}\\
&=&\frac{2}{\pi}\int_0^{2\pi}d\phi\int_0^\infty e^{-3 r^2}\frac{r^{n+m}}{\sqrt{n!m!}}e^{i(m-n)\phi} rdr\\
&=&\delta_{mn}\frac{2}{3^{n+1}},
\end{eqnarray}
where use has been made of the polar coordinates $(r,\phi)$. In fact we can rewrite the above result in the from
\begin{equation}
\langle n|\hat\rho|n\rangle=\frac{\bigl(\frac{1}{2}\bigr)^n}{(1+\frac{1}{2})^{n+1}}.\label{nom1}
\end{equation} 
This reminds us the form of the occupation number representation of the density matrix of a harmonic oscillator in thermal equilibrium at temperature $T$, namely,
\begin{equation}
\hat\rho=\sum_n\frac{\langle n\rangle^n}{(1+\langle n\rangle)^{n+1}}|n\rangle\langle n|\label{nom2}
\end{equation}
where the mean value is given in terms of the natural frequency of the oscillator by
\begin{equation}
\langle n\rangle={\rm tr}(\hat\rho a^\dag a)=\frac{1}{e^{\frac{\hbar\omega}{k_BT}}-1}.\label{nom3}
\end{equation}

Comparing equations~(\ref{nom1}) and (\ref{nom2}), and using equation~(\ref{nom3}), we obtain
\begin{equation}
\langle n\rangle=\frac{1}{2},\quad \frac{\hbar\omega}{k_BT}={\ln 3},\label{boson1}
\end{equation}
meaning that
\begin{equation}
\hat\rho=e^{-\ln 3(a^\dag a+\frac{1}{2})}/Z, \quad Z=\frac{\sqrt{3}}{2}.\label{boson2}
\end{equation}

 Our main result may thus be formulated as follows ($\hbar=1$):
\begin{theorem} Given any well-behaved function $f$, we have:  
\begin{equation}\boxed{
\lim_{N\to\infty}{\rm tr}\Biggl\{\underbrace{\frac{\mathbb I_N}{2^N}}_{\hat\rho_s(T=\infty)}f\Bigl(\frac{\hat S_+}{\sqrt{N}},\frac{\hat S_-}{\sqrt{N}}\Bigr)\Biggr\}=\frac{2}{\sqrt{3}}{\rm tr}\Bigl\{e^{-\ln 3(a^\dag a+\frac{1}{2})}{\mathcal N}f(a^\dag,a)\Bigr\}}\label{result}
\end{equation}
\end{theorem}
We conclude that at infinite temperature and large $N$, the raising and lowering scaled spin operators $\hat S_+/\sqrt{N}$ and $\hat S_-/\sqrt{N}$ behave like the creation and annihilation  operators $a^\dag$ and $a$ of a harmonic oscillator in thermal equilibrium whose frequency and temperature are related by $\hbar\omega/k_BT=\ln 3$. 

\subsubsection*{Worked example} Let us illustrate the above result by considering the  operator
 \begin{equation}
f\Bigl(\frac{\hat S_+}{\sqrt{N}},\frac{\hat S_-}{\sqrt{N}}\Bigr)=(\hat S_+\hat S_-+\hat S_-\hat S_+)^5/N^{5/2}.
\end{equation} 
For $N=2000$ we find that
\begin{equation}
{\rm tr}\Biggl\{\frac{\mathbb I_N}{2^N}f\Bigl(\frac{\hat S_+}{\sqrt{N}},\frac{\hat S_-}{\sqrt{N}}\Bigr)\Biggr\}=119.670.
\end{equation}
On the other hand, 
\begin{equation}
\mathcal N f(a^\dag,a)=2^5 {a^\dag}^5 a^5=32[(a^\dag a)^5-10(a^\dag a)^4+35 (a^\dag a)^3-50 (a^\dag a)^2+24 a^\dag a].
\end{equation}
Taking into account the fact  that
\begin{equation}
\frac{2}{\sqrt{3}}\sum_{n=0}^\infty e^{-\ln3(n+\frac{1}{2})} n^k=\frac{2}{3}{\rm Li}_{-k}\Bigl(\frac{1}{3}\Bigr)
\end{equation}
where ${\rm Li_m(x)}$ denotes the polylogarithmic function, we find that
\begin{equation}
\frac{2}{\sqrt{3}}\sum _{n=0}^\infty 32 e^{-\ln3(n+\frac{1}{2})} (n^5-10 n^4+35 n^3-50 n^2+24 n)=120
\end{equation}
as  should be. 

What about the $z$ component? To determine the equivalent  degree of freedom, let us have a look at the probability density
\begin{equation}
x\mapsto p(x)=\sqrt{\frac{2}{\pi}} e^{-2 x^2}
\end{equation}
and compare it with the Gaussian probability distribution of a harmonic oscillator [see equation~(\ref{imad})]
\begin{eqnarray}
  && x\mapsto |\psi(x)|^2=\frac{1}{\sqrt{2\pi(\Delta x)^2}}\exp\Bigl\{-\frac{(x-\langle x\rangle)^2}{2(\Delta x)^2}\Bigr\},\\ 
&&\langle f(x)\rangle=\int_{-\infty}^{+\infty}f(x)|\psi(x)|^2 dx.
\end{eqnarray} 
We find that 
\begin{equation}
\langle x\rangle=0, \quad \Delta x=\frac{1}{2}.
\end{equation}
For the ground state (that is a minimum uncertainty state), the variance is given in terms of the mass $m$ and the frequency $\tilde\omega$ of the oscillator by $\Delta x=\sqrt{\frac{1}{2m\tilde\omega}}$ (remember that we set $\hbar=1$). Hence
\begin{equation}
m\tilde\omega=2.
\end{equation}
This means that, as $N\to\infty$, the operator $\hat S_z/\sqrt{N}$  behaves like the {\it position variable} of a  harmonic oscillator (different from the above one) in its  Gaussian  ground  state ($T=0$) whose variance is equal to $1/2$. The full spin system is thus mapped onto a bipartite bosonic system, consisting of two independent harmonic oscillators, one of which is in thermal equilibrium, while the other one is at zero temperature. 

Let us apply the above results to the Heisenberg $XY$ Hamiltonian ($\gamma$ is the coupling constant):
\begin{equation}
\hat H=\frac{\gamma}{N}(\hat S_+\hat S_-+\hat S_-\hat S_+).
\end{equation}
The thermal expectation value of the function operator $f(\hat S_+/\sqrt{N},\hat S_-/\sqrt{N})$ is given by
\begin{equation}
\langle f\rangle=\frac{{\rm tr}\ e^{-\frac{\gamma}{k_B T N}(\hat S_+\hat S_-+\hat S_-\hat S_+)} f\Bigl(\frac{\hat S_+}{\sqrt{N}},\frac{\hat S_-}{\sqrt{N}}\Bigr)}{{\rm tr}\ e^{-\frac{\gamma}{k_B T N}(\hat S_+\hat S_-+\hat S_-\hat S_+)} }
\end{equation} 
As $N\to\infty$, equation~(\ref{result}) implies that
\begin{equation}
\langle f \rangle =\frac{{\rm tr} \ e^{-\ln 3(a^\dag a+\frac{1}{2})}{\mathcal N} e^{-\frac{\gamma}{k_B T }(a^\dag a+a a^\dag)} f(a^\dag,a)}{{\rm tr}\ e^{-\ln 3(a^\dag a+\frac{1}{2})}{\mathcal N} e^{-\frac{\gamma}{k_B T }(a^\dag a+ a a^\dag)} }\label{free},
\end{equation}
where $\gamma$ is now given in units of $\hbar$. Taking into account the fact that
\begin{equation}
{\mathcal N} (a^\dag a+a a^\dag)^n= 2^n (a^\dag)^n a^n=2^n\sum_{\ell=0}^n B^n_\ell (a^\dag a)^\ell
\end{equation}
where $B^n_\ell$ denotes Stirling's numbers of the first kind~\cite{14}, we find that
\begin{equation}
{\mathcal N} e^{-\frac{\gamma}{k_B T }(a^\dag a+a a^\dag)}=\sum_{n=0}^\infty\Bigl(\frac{-2\gamma}{k_BT}\Bigr)^n\frac{1}{n!}\sum_{\ell=0}^nB^n_\ell(a^\dag a)^\ell.
\end{equation}
By using the following formula for Stirling's numbers of the first kind:
\begin{equation}
(1+t)^u=\sum_{n=0}^\infty\sum_{k=0}^n B^n_k\frac{t^n}{n!} u^k,
\end{equation}
we deduce that
\begin{equation}
{\mathcal N} e^{-\frac{\gamma}{k_B T }(a^\dag a+a a^\dag)}=(1-2\gamma/(k_BT))^{a^\dag a}=e^{\ln(1-\frac{2\gamma}{k_BT})a^\dag a}.
\end{equation}
From equation~(\ref{free}), we find that the partition function of the corresponding bosonic system is given by
\begin{equation}
Z={\rm tr}\Bigl\{e^{-\ln\bigl(\frac{3}{1-2\gamma/k_B T}\bigr)(a^\dag a+\frac{1}{2})}\Bigr\},
\end{equation}
which is the partition function for a single mode harmonic oscillator in thermal equilibrium. The following conditions should however be satisfied:
\begin{equation}
1>\frac{2\gamma}{k_B T}, \quad 1>-\frac{\gamma}{k_B T}.
\end{equation}
In the case of ferromagnetic interactions, $\gamma<0$,  we find that
\begin{equation}
k_B T>|\gamma|.
\end{equation} 
This gives the lower bound on the temperature above which the bosonization is valid. To proceed further, let us apply equation~(\ref{nova}) to the low excitation sector of the Hamiltonian; one finds
\begin{equation}
\hat H=\gamma(\alpha_0^\dag\alpha_0+h.c)=2\gamma(\alpha_0^\dag\alpha_0+\frac{1}{2}).
\end{equation}
Hence,
\begin{equation}
k_BT>\frac{\hbar\omega_0}{2}, \qquad \hbar\omega_0=2|\gamma|.
\end{equation}
We may thus define an effective temperature for the harmonic oscillator as follows:
\begin{equation}
k_BT_{\rm eff}=\frac{2|\gamma|}{\ln\bigl(\frac{3}{1-2\gamma/k_B T}\bigr)}.
\end{equation}
In the case of antiferromagnetic interactions, $\gamma>0$,  we find that
\begin{equation}
k_B T>2\gamma,
\end{equation} 
meaning that the lower bound is fixed by the condition
\begin{equation}
k_BT>\hbar\omega_0.
\end{equation}
This allows us to introduce an effective temperature that has the same form as that corresponding to the ferromagnetic case, except that
the coupling constant, $\gamma$, is now positive.

Finally, regarding the function $f$, it can be mapped according to
\begin{equation}
f\Biggl(\frac{\hat S_+}{\sqrt{N}},\frac{\hat S_-}{\sqrt{N}}\Biggr )\mapsto e^{-\ln(1-2\gamma/k_BT)a^\dag a} {\mathcal N} e^{-\frac{2\gamma}{k_B T }(a^\dag a+\frac{1}{2})} f(a^\dag,a).
\end{equation}
\section{Discussion and concluding remarks} 
In conclusion we have established a bosonic representation for scaled total spin operators at high temperatures. It should be pointed out that this bosonization scheme does not preserve the temperature, since  the spin system, originally taken at infinite temperature is mapped onto a bosonic system at finite temperature. Equation~(\ref{boson1}) does not determine the exact values of the parameters $\omega$ and $T$, but knowing the value of the fraction is sufficient to fully determine the trace.  This can be physically explained by the fact that the original spin system is fully randomized at infinite temperature; the number of degrees of freedom necessary to fully characterize it is equal to the number of the  generators $\hat S_k$, namely, 3. But for the harmonic oscillator, we have only two generators; the $z$ component is found to be equivalent to the position variable of another harmonic oscillator that is at zero temperature. This  discrepancy  is removed by the value $\ln 3$ of the fraction $\hbar\omega/k_BT$ which is clearly the entropy of a fully mixed three-level system.  We have applied the formalism to the Heisenberg $XY$ model at finite temperature, and found that it is equivalent to a single mode harmonic oscillator in thermal equilibrium which can be assigned an effective temperature  that depends logarithmically on the physical one. Moreover, we have determined the exact rule for mapping any function of the spin operators to the bosonic counter part. The presented results in this paper are a first attempt and further investigation needs to be done to gain more insight into other possible applications.

\end{document}